\documentstyle[12pt]{article}
\def\beq{\begin{equation}}
\def\eeq{\end{equation}}
\def\beqn{\begin{eqnarray}}
\def\eeqn{\end{eqnarray}}
\def\dis{\displaystyle}
\def\l{\left}
\def\r{\right}

\begin{document}

\baselineskip=21pt

\begin{flushright}
{\bf LMU-14/94} \\
{August, 1994}
\end{flushright}

\vspace{0.3cm}

\begin{center}
{\Large\bf Isolating the Penguin-diagram Contribution} \\
{\Large\bf to $CP$ Violation in $B^{0}_{d}$ vs $\bar{B}^{0}_{d}\rightarrow
\pi^{+}\pi^{-}$}
\end{center}

\vskip .6in

\begin{center}
{\bf Zhi-zhong XING}\footnote{E-mail: Xing$@$hep.physik.uni-muenchen.de}\\
{\sl Sektion Physik der Universit${\sl\ddot a}$t M${\sl\ddot u}$nchen,
D-80333 Munich, Germany}
\end{center}

\vspace{3cm}

\begin{abstract}

	A reliable prediction for $CP$ violation in $B^{0}_{d}$ vs $\bar{B}^{0}_{d}
\rightarrow \pi^{+}\pi^{-}$ suffers from the penguin-diagram induced
uncertainty.
With the help of SU(3) relations, we show that both the magnitude and strong
phase shift
of the penguin amplitude can be approximately determined only
from the branching ratios of $B^{0}_{d}\rightarrow \pi^{+}\pi^{-}$,
$B^{0}_{d}\rightarrow K^{+}\pi^{-}$, and $B^{+}_{u}\rightarrow K^{0}\pi^{+}$.

\end{abstract}

\vspace{4cm}

\begin{center}
\end{center}

\newpage

	In addition to $B^{0}_{d}\rightarrow J/\psi K_{S}$,
$B^{0}_{d}\rightarrow \pi^{+}\pi^{-}$
is another very promising process for the study of $CP$ violation in
neutral $B$ decays [1]. However, the penguin-diagram induced effect on
$B^{0}_{d}$ vs $\bar{B}^{0}_{d} \rightarrow \pi^{+}\pi^{-}$ may be quite large
so that a
reliable prediction for the $CP$-violating asymmetry between them is
difficult [2,3]. It has been shown that the penguin contributions to neutral
$B$
decays into $CP$ eigenstates can in principle be rigorously determined with
detailed studies of the time dependence of several correlative decay modes
[4,5].
Long before such measurements are carried out at asymmetric $e^{+}e^{-}$ $B$
factories
or hadron machines, it is very useful at present to pursue an approximate
way to probe the penguin effect on $CP$ violation by using the accessible
(time-independent)
data of $B$-meson decays.

	In Ref. [6] Silva and Wolfenstein have applied SU(3) symmetry to
$B^{0}_{d}\rightarrow
\pi^{+}\pi^{-}$ and $B^{0}_{d}\rightarrow K^{+}\pi^{-}$ as a probe of the
penguin
contribution. Neglecting the final-state interactions which generally produce a
strong
phase difference between the penguin and tree amplitudes,
they found that the magnitude of the penguin amplitude could be approximately
determined
from the ratio of the decay rates of these two processes.
A more general and complete application of SU(3) symmetry to $B$ decays into
two pseudoscalar mesons has recently been given by Gronau et al [7]. With some
triangle relations among the decay amplitudes of $B\rightarrow \pi\pi, \pi K$,
and
$KK$, they explored various possibilities to extract the weak
Cabibbo-Kobayashi-Maskawa
(CKM) phases and the strong final-state phase shifts only through the
time-independent
measurements of decay rates.
In this letter, we shall concentrate on the penguin-induced strong interections
in the SU(3)-related transitions
$B^{0}_{d}\rightarrow \pi^{+}\pi^{-}$,
$B^{0}_{d}\rightarrow K^{+}\pi^{-}$, and $B^{+}_{u}\rightarrow K^{0}\pi^{+}$.
We show that both the relative magnitude and strong phase shift between the
penguin and
tree amplitudes are determinable only if the branching ratios of these decays
are measured.
The signals of direct and indirect $CP$ violation in
$B^{0}_{d}$ vs $\bar{B}^{0}_{d}\rightarrow \pi^{+}\pi^{-}$
can be consequently determined without observing the time dependence of decay
rates.
Using the effective weak Hamiltonian and factorization approximation,
we estimate SU(3) breaking in the tree and penguin amplitudes for decays of the
types
$B\rightarrow \pi\pi$ and $B\rightarrow K\pi$.

%-----------------------------------------------

	We begin with a graphical description for the decay modes in question.
To the lowest-order weak interactions in the standard model,
any $B$ decays into two light mesons can be described using a set of ten
topologically distinct quark diagrams [8]. As shown in Fig. 1,
the diagrams $n$ and $n'$ (with
$n=1,\cdot\cdot\cdot,5$) are different from each other in the final-state
hadronization of the valence quarks. The overall amplitude of a
decay mode $B\rightarrow f$ is expressed in terms of the quark-diagram
amplitudes,
$A_{n}$ and $A_{n'}$, as
\beq
A(B\rightarrow f) \; =\; \dis\sum_{n=1}^{5} C_{n}A_{n}(f) +
\dis\sum_{n'=1}^{5} C_{n'}A_{n'}(f) \; ,
%		(1)
\eeq
where $C_{n}$ and $C_{n'}$ are real coefficients.
Adopting the usual valence-quark conventions for the SU(3) mesons [9], we
obtain from Fig. 1 that
\beqn
A(B^{0}_{d}\rightarrow \pi^{+}\pi^{-}) & = & A_{1}(\pi^{+}\pi^{-})
+A_{2}(\pi^{+}\pi^{-})+A_{4}(\pi^{+}\pi^{-})+A_{5}(\pi^{+}\pi^{-}) \; ,
\nonumber \\
A(B^{+}_{u}\rightarrow K^{0}\pi^{+}) & = &
A_{3}(K^{0}\pi^{+})+A_{4}(K^{0}\pi^{+}) \; , \\
A(B^{0}_{d}\rightarrow K^{+}\pi^{-}) & = &
A_{1}(K^{+}\pi^{-})+A_{4}(K^{+}\pi^{-}) \; . \nonumber
%		(2)
\eeqn

	Subsequently we neglect the amplitudes $A_{2}(\pi^{+}\pi^{-})$,
$A_{5}(\pi^{+}\pi^{-})$, and $A_{3}(K^{0}\pi^{+})$, which have been argued to
be
formfactor- or helicity-suppressed in comparison with those
from the spectator-type diagrams $1$ and $4$ [10]. In addition,
we assume that the
penguin amplitudes $A_{4}(\pi^{+}\pi^{-})$ and $A_{4}(K^{0}\pi^{+})$ (or
$A_{4}(K^{+}\pi^{-})$) are
dominated by the top-quark loop and therefore have the weak phases
$\arg(V^{*}_{tb}V_{td})$ and $\arg(V^{*}_{tb}V_{ts})$, respectively.
The approach to test these two approximations has been suggested in Ref. [7].
Factoring out the weak-interaction part (i.e., the
CKM matrix elements), the decay amplitudes in Eq. (2) are approximately given
by
\beqn
A(B^{0}_{d}\rightarrow \pi^{+}\pi^{-}) & = & A\lambda \left [ r
e^{i\gamma}\tilde{A}_{1}(\pi^{+}\pi^{-})
+ s e^{-i\beta}\tilde{A}_{4}(\pi^{+}\pi^{-}) \right ] \; , \nonumber \\
A(B^{+}_{u}\rightarrow K^{0}\pi^{+}) & = & -A\lambda^{2}
\tilde{A}_{4}(K^{0}\pi^{+}) \; , \\
A(B^{0}_{d}\rightarrow K^{+}\pi^{-}) & = & A\lambda^{2} \left [ r e^{i\gamma}
\tilde{A}_{1}
(K^{+}\pi^{-}) - \tilde{A}_{4}(K^{+}\pi^{-}) \right ] \; , \nonumber
%		(3)
\eeqn
where
\beq
\begin{array}{lll}
r \; = \; \lambda^{2}\sqrt{\rho^{2}+\eta^{2}} \; , &\;\;\;\;\;\; & s \;
= \; \lambda^{2}\sqrt{(1-\rho)^{2}+\eta^{2}} \; , \\
\beta \; = \; \arctan \displaystyle\left (\frac{\eta}{\rho}\right ) \; ,
&  & \gamma \; = \; \arctan \displaystyle\left (\frac{\eta}{1-\rho}\right ) \;
{}.
\end{array}
%		(4)
\eeq
In Eqs. (3) and (4), $A, \lambda, \rho$, and $\eta$ are the Wolfenstein
parameters of the CKM matrix.
The phases $\beta$ and $\gamma$ correspond to two angles of the CKM unitarity
triangle.

	With isospin symmetry, one obtains $\tilde{A}_{4}(K^{0}\pi^{+})
=\tilde{A}_{4}(K^{+}\pi^{-})$. The SU(3) invariance indicates
$\tilde{A}_{1}(\pi^{+}\pi^{-})=\tilde{A}_{1}(K^{+}\pi^{-})$ and
$\tilde{A}_{4}(\pi^{+}\pi^{-})=\tilde{A}_{4}(K^{+}\pi^{-})$. In practice,
one should take into account SU(3) violation for the $B\rightarrow \pi\pi$
and $B\rightarrow K\pi$ transitions. Let us estimate the SU(3)
breaking effect on the above reduced amplitudes using the
factorization approximation. The effective weak Hamiltonian responsible for
the transitions $\bar{b}\rightarrow (\bar{u}u)\bar{q}$ (with $q=d$ or $s$) is
[11]
\begin{equation}
{\cal H}_{\rm eff}(\Delta B=+1)\;=\;\frac{G_{F}}{\sqrt{2}}
\left [V^{*}_{ub}V_{uq}\left (\bar{c}_{1}Q^{q}_{1} + \bar{c}_{2}Q^{q}_{2}
\right )
-V^{*}_{tb}V_{tq}\left (\sum_{i=3}^{6}\bar{c}_{i}Q^{q}_{i}\right )\right ]\; ,
%		(5)
\end{equation}
where $\bar{c}_{i}$ ($i=1,\cdot\cdot\cdot, 6$) are the Wilson coefficients, and
$Q^{q}_{i}$ form an operator basis defined by
\begin{eqnarray}
& Q^{q}_{1} & =\; (\bar{b}_{\alpha}u_{\beta})^{~}_{V-A}
(\bar{u}_{\beta}q_{\alpha})^{~}_{V-A}\; ,\;\;\;\;\;
Q^{q}_{2}\; =\; (\bar{b}u)^{~}_{V-A}(\bar{u}q)^{~}_{V-A}\; ,\nonumber\\
& Q^{q}_{3} & =\; (\bar{b}q)^{~}_{V-A}
\sum_{q^{'}}(\bar{q}^{'}q^{'})^{~}_{V-A}\; ,\;\;\;\;\;
Q^{q}_{4}\; =\; (\bar{b}_{\alpha}q_{\beta})^{~}_{V-A}
\sum_{q^{'}}(\bar{q}^{'}_{\beta}q^{'}_{\alpha})^{~}_{V-A}\; ,\\
& Q^{q}_{5} & =\; (\bar{b}q)^{~}_{V-A}
\sum_{q^{'}}(\bar{q}^{'}q^{'})^{~}_{V+A}\; ,\;\;\;\;\;
Q^{q}_{6}\; =\; (\bar{b}_{\alpha}q_{\beta})^{~}_{V-A}
\sum_{q^{'}}(\bar{q}^{'}_{\beta}q^{'}_{\alpha})^{~}_{V+A}\; .\nonumber
%		(6)
\end{eqnarray}
In the factorization approximation, we obtain
\beq
\begin{array}{lll}
\dis\frac{\tilde{A}_{1}(K^{+}\pi^{-})}{\tilde{A}_{1}(\pi^{+}\pi^{-})}
& = & \dis\frac{\langle
K^{+}\pi^{-}|\bar{c}_{1}Q^{s}_{1}+\bar{c}_{2}Q^{s}_{2}|B^{0}_{d}\rangle }
{\langle
\pi^{+}\pi^{-}|\bar{c}_{1}Q^{d}_{1}+\bar{c}_{2}Q^{d}_{2}|B^{0}_{d}\rangle }
\; =\; \dis\frac{f_{K}}{f_{\pi}} \; , \\
\dis\frac{\tilde{A}_{4}(K^{+}\pi^{-})}{\tilde{A}_{4}(\pi^{+}\pi^{-})}
& = & \dis\frac{\langle K^{+}\pi^{-} \l |\dis\sum_{i=3}^{6}\bar{c}_{i}Q_{i}^{s}
\r |B^{0}_{d}\rangle }
{\langle \pi^{+}\pi^{-} \l |\dis\sum_{i=3}^{6}\bar{c}_{i}Q^{d}_{i} \r
|B^{0}_{d}\rangle }
\; =\; {\cal C} \dis\frac{f_{K}}{f_{\pi}} \; ,
%		(7)
\end{array}
\eeq
where the factor
\beq
{\cal C} \; =\; \frac{\dis\l (\frac{\bar{c}_{3}}{N_{c}}+\bar{c}_{4}\r ) +
\dis\l (\frac{\bar{c}_{5}}{N_{c}}+\bar{c}_{6}\r )
\dis\frac{2m^{2}_{K}}{(m_{u}+m_{s})m_{b}}}
{\dis\l (\frac{\bar{c}_{3}}{N_{c}}+\bar{c}_{4}\r ) + \dis\l
(\frac{\bar{c}_{5}}{N_{c}}+\bar{c}_{6}\r )
\dis\frac{2m^{2}_{\pi}}{(m_{u}+m_{d})m_{b}}}
%		(8)
\eeq
arises from transforming the $(V-A)(V+A)$ currents of $Q^{q}_{5,6}$ into the
$(V-A)(V-A)$ ones. Following the approximate rule of discarding $1/N_{c}$
corrections
in nonleptonic exclusive $B$ decays [12], we estimate $\cal C$ using
$m_{u}=5$ MeV, $m_{d}=10$ MeV, $m_{s}=175$ MeV, $m_{b}=4.8$ GeV,
$\bar{c}_{4}=-0.034$, and $\bar{c}_{6}=-0.042$ [11]. It turns out that ${\cal
C}\approx 1.0$. In addition,
we have $f_{K}/f_{\pi}\approx 1.2$ from the existing experimental data [9].
These two
numbers serve as an illustration of the magnitude of SU(3) violation
in $B^{0}_{d}\rightarrow \pi^{+}\pi^{-}$ and $B^{0}_{d}\rightarrow
K^{+}\pi^{-}$.

%-----------------------------------------------

	The penguin effect on $B^{0}_{d}\rightarrow \pi^{+}\pi^{-}$ can be described
by the modulus and phase of the ratio of $\tilde{A}_{4}(\pi^{+}\pi^{-})$ and
$\tilde{A}_{1}(\pi^{+}\pi^{-})$:
\beq
\chi \; = \; \l
|\frac{\tilde{A}_{4}(\pi^{+}\pi^{-})}{\tilde{A}_{1}(\pi^{+}\pi^{-})}\r |
\; , \;\;\;\;\;\;\; \theta \; = \; \arg \l [
\frac{\tilde{A}_{4}(\pi^{+}\pi^{-})}
{\tilde{A}_{1}(\pi^{+}\pi^{-})} \r ] \; .
%		(9)
\eeq
Using the SU(3) relations given in Eq. (7), we connect $\chi$ and $\theta$ to
the following
two observables:
\beq
R_{1} \; = \; \l |\frac{A(B^{0}_{d}\rightarrow K^{+}\pi^{-})}
{A(B^{+}_{u}\rightarrow K^{0}\pi^{+})} \r |^{2} \; ,
\;\;\;\;\;\;\; R_{2}\; = \; \l |\frac{A(B^{0}_{d}\rightarrow \pi^{+}\pi^{-})}
{A(B^{+}_{u}\rightarrow K^{0}\pi^{+})} \r |^{2} \; .
%		(10)
\eeq
$R_{1}$ and $R_{2}$ can be determined from the branching ratios of the
three decay modes in question. We find
\beq
\begin{array}{lll}
R_{1} & = & 1+\dis\frac{r^{2}}{{\cal C}^{2}\chi^{2}} - \dis\frac{2r}{{\cal
C}\chi}
\cos (\gamma -\theta) \; , \\
R_{2} & = & \dis\frac{s^{2}}{\lambda^{2}{\cal C}^{2}} \l (\dis\frac{f_{\pi}}
{f_{K}}\r )^{2} \l [ 1+\dis\frac{r^{2}}{s^{2}\chi^{2}}
+\dis\frac{2r}{s\chi} \cos (\beta + \gamma -\theta ) \r ] \; .
\end{array}
%		(11)
\eeq
The weak phases $\beta$ and $\gamma$ are related to the real parameters
$r$ and $s$ (or $\rho$ and $\eta$) through Eq. (4). A precise
measurement of $\epsilon^{~}_{K}$ (the $CP$-violating parameter in the
$K^{0}-\bar{K}^{0}$ system) and $B^{0}_{d}-\bar{B}^{0}_{d}$ mixing may fix the
values of $\beta$ and $\gamma$.
Independently these two phases can be measured in some
distinct nonleptonic decays of $B$ mesons. For instance, it is possible to
extract $\beta$ from the $CP$ asymmetry in $B^{0}_{d}$ vs $\bar{B}^{0}_{d}
\rightarrow J/\psi K_{S}$ [1] and $\gamma$ from the decay rates of
$B^{\pm}_{u}\rightarrow K^{\pm}D^{0}, K^{\pm}\bar{D}^{0}$, and $K^{\pm}D_{CP}$
[13].
Assuming $\beta$ and $\gamma$ are determined and $R_{1}$ and $R_{2}$ are
measured, we may use Eq. (11) to probe the strong-interaction parameters
$\chi$ and $\theta$. Keeping only the positive $\chi$, the value of
$\theta$ will be determinable with only a two-fold discrete ambiguity.

	We proceed to discuss the penguin effect on the $CP$-violating
asymmetry in $B^{0}_{d}$ vs $\bar{B}^{0}_{d}\rightarrow \pi^{+}\pi^{-}$.
For either time-dependent or time-integrated measurements, $CP$ violation
is signified by the following two observables [5]:
\beq
T_{\pi^{+}\pi^{-}} \; =\;
\frac{1-|\zeta_{\pi^{+}\pi^{-}}|^{2}}{1+|\zeta_{\pi^{+}\pi^{-}}|^{2}} \; ,
\;\;\;\;\;\;\; T'_{\pi^{+}\pi^{-}} \; =\; \frac{-2{\rm Im}\displaystyle\left
(e^{-2i\beta}\zeta_{\pi^{+}\pi^{-}}
\right )}{1+|\zeta_{\pi^{+}\pi^{-}}|^{2}} \; ,
%		(12)
\eeq
where $\zeta_{\pi^{+}\pi^{-}} = A(\bar{B}^{0}_{d}\rightarrow \pi^{+}\pi^{-})/
A(B^{0}_{d}\rightarrow \pi^{+}\pi^{-})$.
The nonvanishing $T_{\pi^{+}\pi^{-}}$ and $T'_{\pi^{+}\pi^{-}}$ imply direct
$CP$ violation
in the decay amplitude and indirect
$CP$ violation from the interference between decay and
$B^{0}_{d}-\bar{B}^{0}_{d}$
mixing, respectively.
With the help of Eqs. (7) and (11), we obtain
\beq
\begin{array}{rll}
|\zeta_{\pi^{+}\pi^{-}}|^{2} & = & \dis\frac{s^{2}}{\lambda^{2}{\cal
C}^{2}R_{2}} \dis\l (\frac{f_{\pi}}
{f_{K}}\r )^{2} \dis\l [ 1+ \dis\frac{r^{2}}{s^{2}\chi^{2}} + \dis\frac{2r}
{s\chi}\cos (\beta +\gamma +\theta ) \r ] \; , \\
{\rm Im}\displaystyle\left (e^{-2i\beta}\zeta_{\pi^{+}\pi^{-}}\right ) & = & -
\dis\frac{r^{2}}{\lambda^{2}
{\cal C}^{2}\chi^{2}R_{2}} \dis\l ( \dis\frac{f_{\pi}}{f_{K}}\r )^{2} \dis\l [
\sin 2(\beta + \gamma)
+ \dis\frac{2s\chi}{r} \sin(\beta +\gamma) \cos \theta \r ] \; .
\end{array}
%		(13)
\eeq
One can observe that the $\sin 2(\beta +\gamma)$ term in $T'_{\pi^{+}\pi^{-}}$
is
maximally corrected by the penguin contribution in the case of $\theta =0$
or $\pm\pi$.

	For illustration, we give a numerical estimate of the
decay-rate ratios ($R_{1}$ and $R_{2}$) and the $CP$-violating observables
($T_{\pi^{+}\pi^{-}}$ and $T'_{\pi^{+}\pi^{-}}$)
as functions of the penguin-amplitude parameters $\chi$ and $\theta$.
The Wolfenstein parameters are taken as $A=0.86, \lambda=0.22, \rho=0.14$, and
$\eta=0.36$,
which are consistent with the current data on $\epsilon^{~}_{K}$ and
$B^{0}_{d}-\bar{B}^{0}_{d}$ mixing [14]. The results are shown in Figs. 2 and
3.
One observes that $\chi$ and $\theta$ can be well determined from
$R_{1}$ and $R_{2}$ if $\chi$ is in the range\footnote{
Within the standard model, several rough estimates have given $\chi\sim 0.1$
(see, e.g.,
Refs. [2,3]).}
$0.05\leq \chi \leq 0.3$. On the other hand,
a nonvanishing $\theta$ is crucial for direct
$CP$ violation ($T_{\pi^{+}\pi^{-}}$) and has significant influence on indirect
$CP$ violation
($T'_{\pi^{+}\pi^{-}}$) .
It should be noted that in Fig. 3 the values of $T_{\pi^{+}\pi^{-}}$ and
$T'_{\pi^{+}\pi^{-}}$ depend sensitively upon the input of the CKM matrix
parameters.

	The combined branching ratio of $B^{0}_{d}\rightarrow \pi^{+}\pi^{-}$
and $B^{0}_{d}\rightarrow K^{+}\pi^{-}$ has been measured to be
$(2.4^{+0.8}_{-0.7}\pm 0.2)\times 10^{-5}$ [15]. It is most likely that the
decay rates of these
two processes are of the same order [7]. Since the tree amplitude of
$B^{0}_{d}\rightarrow K^{+}\pi^{-}$
is $\lambda^{2}$-suppressed in comparison with the penguin amplitude,
one expects that the latter dominates the decay.
Accordingly the branching ratio of $B^{+}_{u}\rightarrow K^{0}\pi^{+}$
should also be of the order $10^{-5}$. With this level of
decay rates the above three transitions will soon be measurable
at current $e^{+}e^{-}$ colliders or hadron machines. It is therefore
possible to obtain some useful information on $CP$ violation in $B^{0}_{d}$
vs $\bar{B}^{0}_{d}\rightarrow \pi^{+}\pi^{-}$, long before the
time-dependent measurements are carred out at the future asymmetric
$B$ factories.

%--------------------------------------------------------

	One can in principle apply SU(3) symmetry to other
$B\rightarrow \pi\pi$ and $B\rightarrow K\pi$ (or $B\rightarrow KK$)
transitions
in order to probe the weak and strong phases only from measurements
of the decay rates [7]. For example, the penguin effect in
$B^{0}_{d}$ vs $\bar{B}^{0}_{d}\rightarrow \pi^{+}\pi^{-}$ may also be
determined by studying $B^{0}_{d}\rightarrow \pi^{+}\pi^{-}$,
$B^{+}_{u}\rightarrow \pi^{+}\pi^{0}$, and $B^{+}_{u}\rightarrow K^{+}\pi^{0}$.
The tree amplitudes of the latter two modes get non-negligible contributions
from the color-suppressed diagrams $1'$, hence we should take account of
SU(3) breaking in the reduced amplitudes $\tilde{A}_{1'}(\pi^{+}\pi^{0})$
and $\tilde{A}_{1'}(K^{+}\pi^{0})$. In the factorization approximation,
one finds
\beq
\begin{array}{lll}
\dis\frac{\tilde{A}_{1'}(K^{+}\pi^{0})}{\tilde{A}_{1'}(\pi^{+}\pi^{0})}
& = & \dis\frac{\langle
K^{+}\pi^{0}|\bar{c}_{1}Q^{s}_{1}+\bar{c}_{2}Q^{s}_{2}|B^{+}_{u}\rangle }
{\langle
\pi^{+}\pi^{0}|\bar{c}_{1}Q^{d}_{1}+\bar{c}_{2}Q^{d}_{2}|B^{+}_{u}\rangle }
\; \approx \; \dis\frac{F^{BK}_{0}(m^{2}_{\pi})}{F^{B\pi}_{0}(m^{2}_{\pi})} \;
, \\
\dis\frac{\tilde{A}_{1'}(\pi^{+}\pi^{0})}{\tilde{A}_{1}(\pi^{+}\pi^{0})}
& = & \dis\frac{ \dis\l (\bar{c}_{1}+\dis\frac{\bar{c}_{2}}{N_{c}}\r )
\langle \pi^{+} | (\bar{u}d)^{~}_{V-A}|0\rangle \langle \pi^{0} |
(\bar{b}u)^{~}_{V-A}|B^{+}_{u}\rangle }
{\dis\l (\bar{c}_{2}+\dis\frac{\bar{c}_{1}}{N_{c}}\r )
\langle \pi^{0}| (\bar{u}u)^{~}_{V-A}|0\rangle \langle  \pi^{+}
|(\bar{b}d)^{~}_{V-A} |B^{+}_{u}\rangle }
\; \approx \; \dis\frac{\bar{c}_{1}}{\bar{c}_{2}} \; ,
%		(14)
\end{array}
\eeq
where the $1/N_{c}$ corrections have been discarded [12].
Using $F^{BK}_{0}(0)=0.379$ and $F^{B\pi}_{0}(0)=0.333$ [16] as well as
$\bar{c}_{1}=-0.291$ and $\bar{c}_{2}=1.133$ [11], we obtain
$F^{BK}_{0}(m^{2}_{\pi})/F^{B\pi}_{0}(m^{2}_{\pi})$
$\approx 1.1$ and $\bar{c}_{1}/\bar{c}_{2}\approx -0.26$. It is expected that
the magnitudes of
$\tilde{A}_{1'}$ and $\tilde{A}_{4}$ are comparable in either
$B^{+}_{u}\rightarrow \pi^{+}\pi^{0}$
or $B^{+}_{u}\rightarrow K^{+}\pi^{0}$. Our estimates in Eqs. (7) and (14)
indicate
that SU(3) breaking in the factorizable amplitudes $\tilde{A}_{1},
\tilde{A}_{1'}$, and
$\tilde{A}_{4}$ seems to be comparable. If this is true, the SU(3) approach
used here and
in Refs. [6,7] should be quite practical for investigating $CP$ violation and
final-state interactions in decays of the types $B\rightarrow \pi\pi, K\pi$,
and $KK$.

	With the help of SU(3) relations, we have shown that the penguin effect on
$CP$ violation in $B^{0}_{d}$ vs $\bar{B}^{0}_{d}\rightarrow \pi^{+}\pi^{-}$
can be
approximately determined only from measurements of the decay rates of
$B^{0}_{d}\rightarrow \pi^{+}\pi^{-}$, $B^{0}_{d}\rightarrow K^{+}\pi^{-}$, and
$B^{+}_{u}\rightarrow K^{0}\pi^{+}$. These three transitions are becoming
accessible in the experiments of $B$ physics. Some other charmless two-body
decays of $B$ mesons,
if they are favored for experimental observation, are worth studying in a
similar way. \\

	I would like to thank Professor H. Fritzsch for his hospitality
and encouragements. I am also grateful to Professor D. Wyler for calling
my attention to the topic under discussion. I finally acknowledge the
financial support from the Alexander-von-Humboldt Foundation.

\newpage

\small
\begin{figure}
\begin{picture}(400,240)
\put(70,215){\line(1,0){90}}
\put(70,190){\line(1,0){90}}
\put(160,240){\oval(70,15)[l]}
\put(145,247.5){\vector(1,0){2}}
\put(145,232.5){\vector(-1,0){2}}
\put(85,215){\vector(1,0){2}}
\put(145,215){\vector(1,0){2}}
\put(85,190){\vector(-1,0){2}}
\put(145,190){\vector(-1,0){2}}
\multiput(110,215)(3,5){5}{\line(0,1){5}}
\multiput(107,215)(3,5){6}{\line(1,0){3}}
\put(113,165){(1)}
%------------------------(1)
\put(250,240){\line(1,0){90}}
\put(250,190){\line(1,0){90}}
\put(340,215){\oval(70,25)[l]}
\put(265,240){\vector(1,0){2}}
\put(265,190){\vector(-1,0){2}}
\put(325,240){\vector(1,0){2}}
\put(325,190){\vector(-1,0){2}}
\put(325,227.5){\vector(-1,0){2}}
\put(325,202.5){\vector(1,0){2}}
\multiput(290,240)(3,-5){5}{\line(0,-1){5}}
\multiput(287,240)(3,-5){6}{\line(1,0){3}}
\put(293,165){($1'$)}
%------------------------(2)
\put(70,114.5){\line(1,0){90}}
\put(70,65.5){\line(1,0){90}}
\put(160,90){\oval(80,25)[l]}
\put(85,114.5){\vector(1,0){2}}
\put(85,65.5){\vector(-1,0){2}}
\put(145,114.5){\vector(1,0){2}}
\put(145,65.5){\vector(-1,0){2}}
\put(145,102.5){\vector(-1,0){2}}
\put(145,77.5){\vector(1,0){2}}
\multiput(100,108.3)(0,-6){8}{$>$}
\put(113,40){(2)}
%------------------------(3)
\put(250,90){\line(1,0){90}}
\put(250,65){\line(1,0){90}}
\put(340,115){\oval(70,15)[l]}
\put(265,90){\vector(1,0){2}}
\put(265,65){\vector(-1,0){2}}
\put(325,90){\vector(1,0){2}}
\put(325,65){\vector(-1,0){2}}
\put(325,122.5){\vector(1,0){2}}
\put(325,107.5){\vector(-1,0){2}}
\multiput(280,84)(0,-6){4}{$>$}
\put(293,40){($2'$)}
%--------------------------(4)
\put(70,-35){\oval(60,31)[r]}
\put(160,-35){\oval(52,21)[l]}
\put(160,-35){\oval(77,46)[l]}
\put(85,-19.5){\vector(1,0){2}}
\put(85,-50.0){\vector(-1,0){2}}
\put(145,-24.5){\vector(-1,0){2}}
\put(145,-45.5){\vector(1,0){2}}
\put(145,-12){\vector(1,0){2}}
\put(145,-58){\vector(-1,0){2}}
\multiput(99.5,-35)(5.2,0){4}{$\wedge$}
\put(113,-85){(3)}
%------------------------(5)
\put(250,-44.5){\oval(60,31)[r]}
\put(340,-44.5){\oval(68,25)[l]}
\put(340,-10){\oval(68,15)[l]}
\put(265,-29){\vector(1,0){2}}
\put(265,-60){\vector(-1,0){2}}
\put(325,-2.5){\vector(1,0){2}}
\put(325,-17.5){\vector(-1,0){2}}
\put(325,-32){\vector(1,0){2}}
\put(325,-57){\vector(-1,0){2}}
\multiput(279.5,-44.5)(5.1,0){5}{$\wedge$}
\put(293,-85){($3'$)}
%---------------------------(6)
\end{picture}

\vspace{-2cm}
\begin{picture}(400,430)
\put(70,240){\line(1,0){25}}
\put(125,240){\line(1,0){35}}
\put(110,240){\oval(30,36)[b]}
\put(70,190){\line(1,0){90}}
\put(160,215){\oval(60,25)[l]}
\put(145,240){\vector(1,0){2}}
\put(145,227.5){\vector(-1,0){2}}
\put(145,202.5){\vector(1,0){2}}
\put(85,240){\vector(1,0){2}}
\put(85,190){\vector(-1,0){2}}
\put(145,190){\vector(-1,0){2}}
\multiput(94,240)(5,0){6}{$\wedge$}
\put(113,165){(4)}
%------------------------(7)
\put(250,215){\line(1,0){30}}
\put(310,215){\line(1,0){30}}
\put(295,215){\oval(30,36)[b]}
\put(250,190){\line(1,0){90}}
\put(340,240){\oval(80,15)[l]}
\put(265,215){\vector(1,0){2}}
\put(265,190){\vector(-1,0){2}}
\put(325,215){\vector(1,0){2}}
\put(325,190){\vector(-1,0){2}}
\put(325,232.5){\vector(-1,0){2}}
\put(325,247.5){\vector(1,0){2}}
\multiput(279,215)(5,0){6}{$\wedge$}
\put(293,165){($4'$)}
%------------------------(8)
\put(70,90){\oval(84,31)[r]}
\put(160,90){\oval(52,21)[l]}
\put(160,90){\oval(74,46)[l]}
\put(85,105.5){\vector(1,0){2}}
\put(85,74.5){\vector(-1,0){2}}
\put(145,113){\vector(1,0){2}}
\put(145,67){\vector(-1,0){2}}
\put(145,100.5){\vector(-1,0){2}}
\put(145,79.5){\vector(1,0){2}}
\multiput(92,99.5)(0,-6){5}{$>$}
\put(113,40){(5)}
%------------------------(9)
\put(250,90){\oval(84,31)[r]}
\put(340,105){\oval(70,18)[l]}
\put(340,75){\oval(70,18)[l]}
\put(265,105.5){\vector(1,0){2}}
\put(265,74.5){\vector(-1,0){2}}
\put(325,114){\vector(1,0){2}}
\put(325,96){\vector(-1,0){2}}
\put(325,84){\vector(1,0){2}}
\put(325,66){\vector(-1,0){2}}
\multiput(272,99.5)(0,-6){5}{$>$}
\put(293,40){($5'$)}
%------------------------(10)
\end{picture}
\vspace{-0.2cm}
\caption{Possible quark diagrams for a $B$ meson decaying into two light
mesons.}
\end{figure}

\newpage

\begin{figure}
\begin{picture}(400,300)
%--------------------------framework----

\put(100,40){\framebox(240,200)}
\put(53,137){$R_{1}$}
\put(80,103.7){1.}
\put(80,170.4){2.}
\put(80,234){3.}
\put(80,40){0.}
\put(100,20){.0}
\put(175,20){.2}
\put(255,20){.4}
\put(332,20){.6}
\put(216,4){$\chi$}
\put(300,65){(a)}
\put(230,215){\circle{3}}
\put(237,215){\circle{3}}
\put(244,215){\circle{3}}
\put(256,212){$\theta=\pm\pi$}
\put(229,199){\framebox(2,2)}
\put(236,199){\framebox(2,2)}
\put(243,199){\framebox(2,2)}
\put(256,196){$\theta=-\pi/2$}
\put(230,185){\circle{3}}
\put(237,185){\circle{3}}
\put(244,185){\circle{3}}
\put(230,185){\circle*{0.5}}
\put(237,185){\circle*{0.5}}
\put(244,185){\circle*{0.5}}
\put(256,182){$\theta=+\pi/2$}
\put(230,170){\circle*{2.5}}
\put(237,170){\circle*{2.5}}
\put(244,170){\circle*{2.5}}
\put(256,167){$\theta=0$}

%---------------------------background -----
\multiput(100,73.3)(237,0){2}{\line(1,0){3}}
\multiput(100,106.7)(235,0){2}{\line(1,0){5}}
\multiput(100,140)(237,0){2}{\line(1,0){3}}
\multiput(100,173.4)(235,0){2}{\line(1,0){5}}
\multiput(100,206.7)(237,0){2}{\line(1,0){3}}
\multiput(140,40)(0,197){2}{\line(0,1){3}}
\multiput(180,40)(0,195){2}{\line(0,1){5}}
\multiput(220,40)(0,197){2}{\line(0,1){3}}
\multiput(260,40)(0,195){2}{\line(0,1){5}}
\multiput(300,40)(0,197){2}{\line(0,1){3}}

%---------------------------- \theta = 0 ---------
\put(104,110.5){\circle*{2.5}}
\put(104.25,97.3){\circle*{2.5}}
\put(104.6,85){\circle*{2.5}}
\put(105,72.4){\circle*{2.5}}
\put(105.5,63.2){\circle*{2.5}}
\put(106,57.4){\circle*{2.5}}
\put(108,50.3){\circle*{2.5}}
\put(112,56.1){\circle*{2.5}}
\put(116,63.9){\circle*{2.5}}
\put(120,70.1){\circle*{2.5}}
\put(124,74.9){\circle*{2.5}}
\put(128,78.7){\circle*{2.5}}
\put(132,81.6){\circle*{2.5}}
\put(136,84.1){\circle*{2.5}}
\put(140,86.1){\circle*{2.5}}
\put(145,88.1){\circle*{2.5}}
\put(150,89.8){\circle*{2.5}}
\put(160,92.4){\circle*{2.5}}
\put(170,94.3){\circle*{2.5}}
\put(180,95.8){\circle*{2.5}}
\put(190,96.7){\circle*{2.5}}
\put(200,97.57){\circle*{2.5}}
\put(210,98.45){\circle*{2.5}}
\put(220,99.3){\circle*{2.5}}
\put(230,99.75){\circle*{2.5}}
\put(240,100.2){\circle*{2.5}}
\put(250,100.65){\circle*{2.5}}
\put(260,101.1){\circle*{2.5}}
\put(270,101.35){\circle*{2.5}}
\put(280,101.6){\circle*{2.5}}
\put(290,101.85){\circle*{2.5}}
\put(300,102.1){\circle*{2.5}}
\put(310,102.3){\circle*{2.5}}
\put(320,102.5){\circle*{2.5}}
\put(330,102.7){\circle*{2.5}}
\put(340,102.9){\circle*{2.5}}

%-------------------------- \theta = \pi ---------
\put(110,235.7){\circle{2.5}}
\put(111,220.9){\circle{2.5}}
\put(112,209){\circle{2.5}}
\put(113.3,196.7){\circle{2.5}}
\put(114.7,186.3){\circle{2.5}}
%\put(116,178.6){\circle{2.5}}
\put(116.4,176.5){\circle{2.5}}
\put(118,169.1){\circle{2.5}}
%\put(120,161.9){\circle{2.5}}
\put(120.5,160.4){\circle{2.5}}
\put(124,151.4){\circle{2.5}}
\put(128,144.2){\circle{2.5}}
\put(132,139){\circle{2.5}}
\put(136,135){\circle{2.5}}
\put(140,131.9){\circle{2.5}}
\put(145,129){\circle{2.5}}
\put(150,126.5){\circle{2.5}}
\put(160,123){\circle{2.5}}
\put(170,120.5){\circle{2.5}}
\put(180,118.7){\circle{2.5}}
\put(190,117.3){\circle{2.5}}
\put(200,116.2){\circle{2.5}}
\put(210,115.3){\circle{2.5}}
\put(220,114.5){\circle{2.5}}
\put(230,113.9){\circle{2.5}}
\put(240,113.4){\circle{2.5}}
\put(250,112.9){\circle{2.5}}
\put(260,112.5){\circle{2.5}}
\put(270,112.1){\circle{2.5}}
\put(280,111.8){\circle{2.5}}
\put(290,111.6){\circle{2.5}}
\put(300,111.3){\circle{2.5}}
\put(310,111.1){\circle{2.5}}
\put(320,110.9){\circle{2.5}}
\put(330,110.7){\circle{2.5}}
\put(340,110.5){\circle{2.5}}

%---------------------------- \theta = \pi/2 ---------
\put(104,240){\circle{3}}
\put(104,240){\circle*{.5}}
\put(104.2,225.4){\circle{3}}
\put(104.2,225.4){\circle*{.5}}
\put(104.4,210.8){\circle{3}}
\put(104.4,210.8){\circle*{.5}}
\put(104.7,192.7){\circle{3}}
\put(104.7,192.7){\circle*{.5}}
\put(105,178){\circle{3}}
\put(105,178){\circle*{.5}}
\put(105.5,159.2){\circle{3}}
\put(105.5,159.2){\circle*{.5}}
\put(106,143.4){\circle{3}}
\put(106,143.4){\circle*{.5}}
\put(107,127.2){\circle{3}}
\put(107,127.2){\circle*{.5}}
\put(108,116.3){\circle{3}}
\put(108,116.3){\circle*{.5}}
\put(109.5,106.5){\circle{3}}
\put(109.5,106.5){\circle*{.5}}
\put(112,100.1){\circle{3}}
\put(112,100.1){\circle*{.5}}
\put(120,96.5){\circle{3}}
\put(120,96.5){\circle*{.5}}
\put(128,97.5){\circle{3}}
\put(128,97.5){\circle*{.5}}
\put(134,98.4){\circle{3}}
\put(134,98.4){\circle*{.5}}
\put(140,99.3){\circle{3}}
\put(140,99.3){\circle*{.5}}
\put(150,100.4){\circle{3}}
\put(150,100.4){\circle*{.5}}
\put(160,101.1){\circle{3}}
\put(160,101.1){\circle*{.5}}
\put(170,101.7){\circle{3}}
\put(170,101.7){\circle*{.5}}
\put(180,102.4){\circle{3}}
\put(180,102.4){\circle*{.5}}
\put(190,102.8){\circle{3}}
\put(190,102.8){\circle*{.5}}
\put(200,103.1){\circle{3}}
\put(200,103.1){\circle*{.5}}
\put(210,103.4){\circle{3}}
\put(210,103.4){\circle*{.5}}
\put(220,103.7){\circle{3}}
\put(220,103.7){\circle*{.5}}
\put(230,103.9){\circle{3}}
\put(230,103.9){\circle*{.5}}
\put(240,104.1){\circle{3}}
\put(240,104.1){\circle*{.5}}
\put(250,104.3){\circle{3}}
\put(250,104.3){\circle*{.5}}
\put(260,104.4){\circle{3}}
\put(260,104.4){\circle*{.5}}
\put(270,104.5){\circle{3}}
\put(270,104.5){\circle*{.5}}
\put(280,104.6){\circle{3}}
\put(280,104.6){\circle*{.5}}
\put(290,104.7){\circle{3}}
\put(290,104.7){\circle*{.5}}
\put(300,104.8){\circle{3}}
\put(300,104.8){\circle*{.5}}
\put(310,104.9){\circle{3}}
\put(310,104.9){\circle*{.5}}
\put(320,105){\circle{3}}
\put(320,105){\circle*{.5}}
\put(330,105){\circle{3}}
\put(330,105){\circle*{.5}}
\put(340,105.1){\circle{3}}
\put(340,105.1){\circle*{.5}}

%-------------------------- \theta = - \pi/2 ---------
%--------- Both x and y have been reduced 1 unit to make the framebox central
%%---------

\put(106,237.4){\framebox(2,2)}
\put(106.5,223.4){\framebox(2,2)}
\put(107,212.6){\framebox(2,2)}
\put(108,195){\framebox(2,2)}
\put(109,181.9){\framebox(2,2)}
\put(110,171.9){\framebox(2,2)}
\put(111,164){\framebox(2,2)}
\put(113,152.5){\framebox(2,2)}
\put(115,144.6){\framebox(2,2)}
\put(119,134.5){\framebox(2,2)}
\put(123,128.4){\framebox(2,2)}
\put(127,124.3){\framebox(2,2)}
\put(133,120.3){\framebox(2,2)}
\put(139,116.7){\framebox(2,2)}
\put(149,114.9){\framebox(2,2)}
\put(159,113.2){\framebox(2,2)}
\put(169,112){\framebox(2,2)}
\put(179,111.1){\framebox(2,2)}
\put(189,110.5){\framebox(2,2)}
\put(199,109.9){\framebox(2,2)}
\put(209,109.5){\framebox(2,2)}
\put(219,109.1){\framebox(2,2)}
\put(229,109.35){\framebox(2,2)}
\put(239,109.6){\framebox(2,2)}
\put(249,108.9){\framebox(2,2)}
\put(259,108.2){\framebox(2,2)}
\put(269,108){\framebox(2,2)}
\put(279,107.9){\framebox(2,2)}
\put(289,107.8){\framebox(2,2)}
\put(299,107.7){\framebox(2,2)}
\put(309,107.6){\framebox(2,2)}
\put(319,107.5){\framebox(2,2)}
\put(329,107.45){\framebox(2,2)}
\put(339,107.4){\framebox(2,2)}

\end{picture}
%*************************Fig.1(a)

\begin{picture}(400,300)
%--------------------------framework----

\put(100,40){\framebox(240,200)}
\put(53,137){$R_{2}$}
\put(80,103.7){1.}
\put(80,170.4){2.}
\put(80,234){3.}
\put(80,40){0.}
\put(100,20){.0}
\put(175,20){.2}
\put(255,20){.4}
\put(332,20){.6}
\put(216,4){$\chi$}
\put(300,65){(b)}
\put(230,215){\circle{3}}
\put(237,215){\circle{3}}
\put(244,215){\circle{3}}
\put(256,212){$\theta=\pm\pi$}
\put(229,199){\framebox(2,2)}
\put(236,199){\framebox(2,2)}
\put(243,199){\framebox(2,2)}
\put(256,196){$\theta=-\pi/2$}
\put(230,185){\circle{3}}
\put(237,185){\circle{3}}
\put(244,185){\circle{3}}
\put(230,185){\circle*{0.5}}
\put(237,185){\circle*{0.5}}
\put(244,185){\circle*{0.5}}
\put(256,182){$\theta=+\pi/2$}
\put(230,170){\circle*{2.5}}
\put(237,170){\circle*{2.5}}
\put(244,170){\circle*{2.5}}
\put(256,167){$\theta=0$}

%---------------------------background -----
\multiput(100,73.3)(237,0){2}{\line(1,0){3}}
\multiput(100,106.7)(235,0){2}{\line(1,0){5}}
\multiput(100,140)(237,0){2}{\line(1,0){3}}
\multiput(100,173.4)(235,0){2}{\line(1,0){5}}
\multiput(100,206.7)(237,0){2}{\line(1,0){3}}
\multiput(140,40)(0,197){2}{\line(0,1){3}}
\multiput(180,40)(0,195){2}{\line(0,1){5}}
\multiput(220,40)(0,197){2}{\line(0,1){3}}
\multiput(260,40)(0,195){2}{\line(0,1){5}}
\multiput(300,40)(0,197){2}{\line(0,1){3}}

%---------------------------- \theta = 0 ---------

\put(117,225.6){\circle*{2.5}}
\put(118,205.7){\circle*{2.5}}
\put(119,188.8){\circle*{2.5}}
\put(120,172.5){\circle*{2.5}}
\put(122,151.4){\circle*{2.5}}
\put(124,133.8){\circle*{2.5}}
\put(126,120.){\circle*{2.5}}
\put(128.8,107.1){\circle*{2.5}}
\put(132,93.5){\circle*{2.5}}
\put(136,82.6){\circle*{2.5}}
\put(140,74.9){\circle*{2.5}}
\put(145,68){\circle*{2.5}}
\put(151,61){\circle*{2.5}}
\put(160,56.5){\circle*{2.5}}
\put(170,52.3){\circle*{2.5}}
\put(180,50){\circle*{2.5}}
\put(190,48.3){\circle*{2.5}}
\put(200,47.1){\circle*{2.5}}
\put(210,46.2){\circle*{2.5}}
\put(220,45.5){\circle*{2.5}}
\put(230,45){\circle*{2.5}}
\put(240,44.5){\circle*{2.5}}
\put(250,44.2){\circle*{2.5}}
\put(260,43.9){\circle*{2.5}}
\put(270,43.7){\circle*{2.5}}
\put(280,43.5){\circle*{2.5}}
\put(290,43.3){\circle*{2.5}}
\put(300,43.2){\circle*{2.5}}
\put(310,43.1){\circle*{2.5}}
\put(320,43){\circle*{2.5}}
\put(330,42.9){\circle*{2.5}}
\put(340,42.8){\circle*{2.5}}

%-------------------------- \theta = \pi ---------

\put(117,227.8){\circle{2.5}}
\put(118,207.4){\circle{2.5}}
\put(119,190.8){\circle{2.5}}
\put(120.5,175){\circle{2.5}}
\put(122,153.1){\circle{2.5}}
\put(124,135.4){\circle{2.5}}
\put(126,121.3){\circle{2.5}}
\put(128.8,108.7){\circle{2.5}}
\put(132,94.7){\circle{2.5}}
\put(136,83.7){\circle{2.5}}
\put(140,75.8){\circle{2.5}}
\put(145,69.1){\circle{2.5}}
\put(151,61.8){\circle{2.5}}
\put(160,57.4){\circle{2.5}}
\put(170,53.1){\circle{2.5}}
\put(180,50.5){\circle{2.5}}
\put(190,48.7){\circle{2.5}}
\put(200,47.5){\circle{2.5}}
\put(210,46.5){\circle{2.5}}
\put(220,45.8){\circle{2.5}}
\put(230,45.3){\circle{2.5}}
\put(240,44.8){\circle{2.5}}
\put(250,44.5){\circle{2.5}}
\put(260,44.1){\circle{2.5}}
\put(270,43.9){\circle{2.5}}
\put(280,43.7){\circle{2.5}}
\put(290,43.55){\circle{2.5}}
\put(300,43.4){\circle{2.5}}
\put(310,43.25){\circle{2.5}}
\put(320,43.1){\circle{2.5}}
\put(330,43.05){\circle{2.5}}
\put(340,43){\circle{2.5}}

%---------------------------- \theta = \pi/2 ---------

\put(119,223.8){\circle{3}}
\put(119,223.8){\circle*{.5}}
\put(120.2,202.1){\circle{3}}
\put(120.2,202.1){\circle*{.5}}
\put(122,181.6){\circle{3}}
\put(122,181.6){\circle*{.5}}
\put(124,161.5){\circle{3}}
\put(124,161.5){\circle*{.5}}
\put(126,145.7){\circle{3}}
\put(126,145.7){\circle*{.5}}
\put(128.5,132.1){\circle{3}}
\put(128.5,132.1){\circle*{.5}}
\put(131,118.3){\circle{3}}
\put(131,118.3){\circle*{.5}}
\put(135,106){\circle{3}}
\put(135,106){\circle*{.5}}
\put(140,91.5){\circle{3}}
\put(140,91.5){\circle*{.5}}
\put(150,76.2){\circle{3}}
\put(150,76.2){\circle*{.5}}
\put(160,67.5){\circle{3}}
\put(160,67.5){\circle*{.5}}
\put(170,62.1){\circle{3}}
\put(170,62.1){\circle*{.5}}
\put(180,58.4){\circle{3}}
\put(180,58.4){\circle*{.5}}
\put(190,55.7){\circle{3}}
\put(190,55.7){\circle*{.5}}
\put(200,53.7){\circle{3}}
\put(200,53.7){\circle*{.5}}
\put(210,52.2){\circle{3}}
\put(210,52.2){\circle*{.5}}
\put(220,51){\circle{3}}
\put(220,51){\circle*{.5}}
\put(230,50.2){\circle{3}}
\put(230,50.2){\circle*{.5}}
\put(240,49.3){\circle{3}}
\put(240,49.3){\circle*{.5}}
\put(250,48.7){\circle{3}}
\put(250,48.7){\circle*{.5}}
\put(260,48.1){\circle{3}}
\put(260,48.1){\circle*{.5}}
\put(270,47.7){\circle{3}}
\put(270,47.7){\circle*{.5}}
\put(280,47.2){\circle{3}}
\put(280,47.2){\circle*{.5}}
\put(290,46.8){\circle{3}}
\put(290,46.8){\circle*{.5}}
\put(300,46.5){\circle{3}}
\put(300,46.5){\circle*{.5}}
\put(310,46.3){\circle{3}}
\put(310,46.3){\circle*{.5}}
\put(320,46){\circle{3}}
\put(320,46){\circle*{.5}}
\put(330,45.8){\circle{3}}
\put(330,45.8){\circle*{.5}}
\put(340,45.6){\circle{3}}
\put(340,45.6){\circle*{.5}}

%-------------------------- \theta = - \pi/2 ---------
%--------- Both x and y have been reduced 1 unit to make the framebox central
%%---------

\put(114,235.2){\framebox(2,2)}
\put(114.7,210){\framebox(2,2)}
\put(116,187.7){\framebox(2,2)}
\put(117,169.8){\framebox(2,2)}
\put(118,154.8){\framebox(2,2)}
\put(119.5,139.1){\framebox(2,2)}
\put(121,121.9){\framebox(2,2)}
\put(124,102.5){\framebox(2,2)}
\put(127,86){\framebox(2,2)}
\put(133,71.9){\framebox(2,2)}
\put(139,58.2){\framebox(2,2)}
\put(149,49.4){\framebox(2,2)}
\put(159,45){\framebox(2,2)}
\put(169,42.6){\framebox(2,2)}
\put(179,41.2){\framebox(2,2)}
\put(189,40.5){\framebox(2,2)}
\put(199,39.8){\framebox(2,2)}
\put(209,39.6){\framebox(2,2)}
\put(219,39.3){\framebox(2,2)}
\put(229,39.25){\framebox(2,2)}
\put(239,39.2){\framebox(2,2)}
\put(249,39.1){\framebox(2,2)}
\put(259,39){\framebox(2,2)}
\put(269,39){\framebox(2,2)}
\put(279,39){\framebox(2,2)}
\put(289,39){\framebox(2,2)}
\put(299,39){\framebox(2,2)}
\put(309,39.05){\framebox(2,2)}
\put(319,39.1){\framebox(2,2)}
\put(329,39.15){\framebox(2,2)}
\put(339,39.2){\framebox(2,2)}

\end{picture}
%================= Fig. (b) =========

\caption{The ratios of decay rates $R_{1}$ and $R_{2}$ as functions of the
penguin-amplitude parameters $\chi$ and $\theta$.}
\end{figure}

\newpage

\begin{figure}
\begin{picture}(400,300)
%--------------------------framework----

\put(100,40){\framebox(240,200)}
\put(42,137){$T_{\pi^{+}\pi^{-}}$}
\put(75,40){$-1.$}
\put(75,86){$-.5$}
\put(80,136){.0}
\put(75,186){$+.5$}
\put(75,234){$+1.$}
\put(100,20){$-\pi$}
\put(150,20){$-\pi/2$}
\put(217,20){0}
\put(268,20){$+\pi/2$}
\put(327,20){$+\pi$}
\put(216,0){$\theta$}
\put(300,65){(a)}
\put(114,224){\line(1,0){2.5}}
\put(121,224){\line(1,0){2.5}}
\put(128,224){\line(1,0){2.5}}
\put(141,222){$\chi=.0$}
\put(115,207){\circle*{3}}
\put(122,207){\circle*{3}}
\put(129,207){\circle*{3}}
\put(141,205){$\chi=.1$}
\put(115,192){\circle{3}}
\put(122,192){\circle{3}}
\put(129,192){\circle{3}}
\put(141,190){$\chi=.3$}

%---------------------------background -----
\multiput(100,90)(235,0){2}{\line(1,0){5}}
\multiput(100,140)(235,0){2}{\line(1,0){5}}
\multiput(100,190)(235,0){2}{\line(1,0){5}}
\multiput(100,215)(237,0){2}{\line(1,0){3}}
\multiput(100,165)(237,0){2}{\line(1,0){3}}
\multiput(100,115)(237,0){2}{\line(1,0){3}}
\multiput(100,65)(237,0){2}{\line(1,0){3}}
\multiput(130,40)(0,197){2}{\line(0,1){3}}
\multiput(160,40)(0,195){2}{\line(0,1){5}}
\multiput(190,40)(0,197){2}{\line(0,1){3}}
\multiput(220,40)(0,195){2}{\line(0,1){5}}
\multiput(250,40)(0,197){2}{\line(0,1){3}}
\multiput(280,40)(0,195){2}{\line(0,1){5}}
\multiput(310,40)(0,197){2}{\line(0,1){3}}
\multiput(100,140)(5,0){48}{\line(1,0){2.5}}
%---------------------------- \chi = 0.1 ---------
\put(100,140){\circle*{3}}
\put(105,134){\circle*{3}}
\put(110,128.3){\circle*{3}}
\put(115,123){\circle*{3}}
\put(120,117.4){\circle*{3}}
\put(125,113){\circle*{3}}
\put(130,108){\circle*{3}}
\put(135,104){\circle*{3}}
\put(140,100.7){\circle*{3}}
\put(145,98){\circle*{3}}
\put(150,96){\circle*{3}}
\put(155,95.1){\circle*{3}}
\put(160,94.3){\circle*{3}}
\put(165,95){\circle*{3}}
\put(170,95.7){\circle*{3}}
\put(175,98){\circle*{3}}
\put(180,100.2){\circle*{3}}
\put(185,103.8){\circle*{3}}
\put(190,107.4){\circle*{3}}
\put(195,112){\circle*{3}}
\put(200,116.9){\circle*{3}}
\put(205,122.4){\circle*{3}}
\put(210,128){\circle*{3}}
\put(215,134){\circle*{3}}
\put(220,140){\circle*{3}}
\put(225,146){\circle*{3}}
\put(230,152){\circle*{3}}
\put(235,157.6){\circle*{3}}
\put(240,163.2){\circle*{3}}
\put(245,167.9){\circle*{3}}
\put(250,172.6){\circle*{3}}
\put(255,176.2){\circle*{3}}
\put(260,179.8){\circle*{3}}
\put(265,181.5){\circle*{3}}
\put(270,184.3){\circle*{3}}
\put(275,185){\circle*{3}}
\put(280,185.7){\circle*{3}}
\put(285,184.8){\circle*{3}}
\put(290,184){\circle*{3}}
\put(295,181.2){\circle*{3}}
\put(300,179.3){\circle*{3}}
\put(305,175.6){\circle*{3}}
\put(310,172){\circle*{3}}
\put(315,167.3){\circle*{3}}
\put(320,162.6){\circle*{3}}
\put(325,157.2){\circle*{3}}
\put(330,151.7){\circle*{3}}
\put(335,146){\circle*{3}}
\put(340,140){\circle*{3}}
%---------------------------- \chi = 0.3 ---------
\put(100,140){\circle{3}}
\put(105,128){\circle{3}}
\put(110,116){\circle{3}}
\put(115,105.8){\circle{3}}
\put(120,93.6){\circle{3}}
\put(125,84){\circle{3}}
\put(130,74){\circle{3}}
\put(135,66.4){\circle{3}}
\put(140,58.8){\circle{3}}
\put(145,53.8){\circle{3}}
\put(150,48.8){\circle{3}}
\put(155,46.9){\circle{3}}
\put(160,44.9){\circle{3}}
\put(165,46.2){\circle{3}}
\put(170,47.5){\circle{3}}
\put(175,52){\circle{3}}
\put(180,56.5){\circle{3}}
\put(185,64){\circle{3}}
\put(190,71.4){\circle{3}}
\put(195,81){\circle{3}}
\put(200,91.2){\circle{3}}
\put(205,103){\circle{3}}
\put(210,114.7){\circle{3}}
\put(215,127){\circle{3}}
\put(220,140){\circle{3}}
\put(225,147.7){\circle{3}}
\put(230,165.3){\circle{3}}
\put(235,177){\circle{3}}
\put(240,188.8){\circle{3}}
\put(245,198.8){\circle{3}}
\put(250,208.6){\circle{3}}
\put(255,216){\circle{3}}
\put(260,223.5){\circle{3}}
\put(265,228.2){\circle{3}}
\put(270,232.5){\circle{3}}
\put(275,233.8){\circle{3}}
\put(280,235.1){\circle{3}}
\put(285,233){\circle{3}}
\put(290,231.2){\circle{3}}
\put(295,226){\circle{3}}
\put(300,221.2){\circle{3}}
\put(305,213.5){\circle{3}}
\put(310,206){\circle{3}}
\put(315,196){\circle{3}}
\put(320,186.4){\circle{3}}
\put(325,175.2){\circle{3}}
\put(330,164){\circle{3}}
\put(335,152){\circle{3}}
\put(340,140){\circle{3}}
\end{picture}
%*************************Fig.1(a)

\begin{picture}(400,300)
%--------------------------framework----
\put(100,40){\framebox(240,200)}
\put(42,137){$T'_{\pi^{+}\pi^{-}}$}
\put(75,40){$-1.$}
\put(75,86){$-.5$}
\put(80,136){.0}
\put(75,186){$+.5$}
\put(75,234){$+1.$}
\put(100,20){$-\pi$}
\put(150,20){$-\pi/2$}
\put(217,20){0}
\put(268,20){$+\pi/2$}
\put(327,20){$+\pi$}
\put(216,0){$\theta$}
\put(300,212){(b)}
\put(114,224){\line(1,0){2.5}}
\put(121,224){\line(1,0){2.5}}
\put(128,224){\line(1,0){2.5}}
\put(141,222){$\chi=.0$}
\put(115,207){\circle*{3}}
\put(122,207){\circle*{3}}
\put(129,207){\circle*{3}}
\put(141,205){$\chi=.1$}
\put(115,192){\circle{3}}
\put(122,192){\circle{3}}
\put(129,192){\circle{3}}
\put(141,190){$\chi=.3$}

%---------------------------background -----
\multiput(100,90)(235,0){2}{\line(1,0){5}}
\multiput(100,140)(235,0){2}{\line(1,0){5}}
\multiput(100,190)(235,0){2}{\line(1,0){5}}
\multiput(100,215)(237,0){2}{\line(1,0){3}}
\multiput(100,165)(237,0){2}{\line(1,0){3}}
\multiput(100,115)(237,0){2}{\line(1,0){3}}
\multiput(100,65)(237,0){2}{\line(1,0){3}}
\multiput(130,40)(0,197){2}{\line(0,1){3}}
\multiput(160,40)(0,195){2}{\line(0,1){5}}
\multiput(190,40)(0,197){2}{\line(0,1){3}}
\multiput(220,40)(0,195){2}{\line(0,1){5}}
\multiput(250,40)(0,197){2}{\line(0,1){3}}
\multiput(280,40)(0,195){2}{\line(0,1){5}}
\multiput(310,40)(0,197){2}{\line(0,1){3}}
\multiput(100,135.4)(5,0){48}{\line(1,0){2.5}}
%---------------------------- \chi = 0.1 ---------
\put(100,89.6){\circle*{3}}
\put(105,90.3){\circle*{3}}
\put(110,91.1){\circle*{3}}
\put(115,93.4){\circle*{3}}
\put(120,95.6){\circle*{3}}
\put(125,99.1){\circle*{3}}
\put(130,102.6){\circle*{3}}
\put(135,107.2){\circle*{3}}
\put(140,111.8){\circle*{3}}
\put(145,117.3){\circle*{3}}
\put(150,122.8){\circle*{3}}
\put(155,129.1){\circle*{3}}
\put(160,135.4){\circle*{3}}
\put(165,141){\circle*{3}}
\put(170,146.3){\circle*{3}}
\put(175,152){\circle*{3}}
\put(180,157.4){\circle*{3}}
\put(185,162){\circle*{3}}
\put(190,166.9){\circle*{3}}
\put(195,170.6){\circle*{3}}
\put(200,174.4){\circle*{3}}
\put(205,176.7){\circle*{3}}
\put(210,179){\circle*{3}}
\put(215,179.8){\circle*{3}}
\put(220,180.6){\circle*{3}}
\put(225,179.8){\circle*{3}}
\put(230,179){\circle*{3}}
\put(235,176.7){\circle*{3}}
\put(240,174.4){\circle*{3}}
\put(245,170.6){\circle*{3}}
\put(250,166.9){\circle*{3}}
\put(255,162){\circle*{3}}
\put(260,157.4){\circle*{3}}
\put(265,152){\circle*{3}}
\put(270,146.3){\circle*{3}}
\put(275,141){\circle*{3}}
\put(280,135.4){\circle*{3}}
\put(285,129.1){\circle*{3}}
\put(290,122.8){\circle*{3}}
\put(295,117.3){\circle*{3}}
\put(300,111.8){\circle*{3}}
\put(305,107.2){\circle*{3}}
\put(310,102.6){\circle*{3}}
\put(315,99.1){\circle*{3}}
\put(320,95.6){\circle*{3}}
\put(325,93.4){\circle*{3}}
\put(330,91.1){\circle*{3}}
\put(335,90.3){\circle*{3}}
\put(340,89.6){\circle*{3}}
%---------------------------- \chi = 0.3 ---------
\put(100,44){\circle{3}}
\put(105,45.5){\circle{3}}
\put(110,47.1){\circle{3}}
\put(115,52.5){\circle{3}}
\put(120,56){\circle{3}}
\put(125,63){\circle{3}}
\put(130,70.5){\circle{3}}
\put(135,80){\circle{3}}
\put(140,89.4){\circle{3}}
\put(145,100.5){\circle{3}}
\put(150,111.8){\circle{3}}
\put(155,123.6){\circle{3}}
\put(160,135.4){\circle{3}}
\put(165,148){\circle{3}}
\put(170,160.9){\circle{3}}
\put(175,172.4){\circle{3}}
\put(180,184.2){\circle{3}}
\put(185,194.3){\circle{3}}
\put(190,204.5){\circle{3}}
\put(195,212.3){\circle{3}}
\put(200,220.2){\circle{3}}
\put(205,225.2){\circle{3}}
\put(210,230.2){\circle{3}}
\put(215,231.9){\circle{3}}
\put(220,233.6){\circle{3}}
\put(225,231.9){\circle{3}}
\put(230,230.2){\circle{3}}
\put(235,225.2){\circle{3}}
\put(240,220.2){\circle{3}}
\put(245,212.3){\circle{3}}
\put(250,204.5){\circle{3}}
\put(255,194.3){\circle{3}}
\put(260,184.2){\circle{3}}
\put(265,172.4){\circle{3}}
\put(270,160.9){\circle{3}}
\put(275,148){\circle{3}}
\put(280,135.4){\circle{3}}
\put(285,123.6){\circle{3}}
\put(290,111.8){\circle{3}}
\put(295,100.5){\circle{3}}
\put(300,89.4){\circle{3}}
\put(305,80){\circle{3}}
\put(310,70.5){\circle{3}}
\put(315,63){\circle{3}}
\put(320,56){\circle{3}}
\put(325,52.5){\circle{3}}
\put(330,47.1){\circle{3}}
\put(335,45.5){\circle{3}}
\put(340,44){\circle{3}}
\end{picture}
%*************************Fig.1(b)
\caption{The $CP$-violating observables $T_{\pi^{+}\pi^{-}}$ and
$T'_{\pi^{+}\pi^{-}}$ as functions of the
penguin-amplitude parameters $\chi$ and $\theta$.}
\end{figure}


\newpage

\baselineskip=16pt

\begin{thebibliography}{99}
\bibitem{1} I.I. Bigi and A.I. Sanda, Nucl. Phys. B193 (1981) 85; B281 (1987)
41.

\bibitem{2} M. Gronau, Phys. Rev. Lett. 63 (1989) 1451; B. Grinstein, Phys.
Lett.
B229 (1989) 280; D. London and R. Peccei, Phys. Lett. B223 (1989) 257;
L.L. Chau and H.Y. Cheng, Phys. Rev. Lett. 59 (1987) 958;
M.B. Gavela et al., Phys. Lett. B154 (1985) 425.

\bibitem{3} D. Du and Z.Z. Xing, Phys. Lett. B280 (1992) 292;
M. Gronau, Phys. Lett. B300 (1993) 163; A. Deandrea et al., Phys. Lett. B320
(1994) 170;
T. Hayashi, M. Matsuda, and M. Tanimoto, Phys. Lett. B323 (1994) 78.

\bibitem{4} M. Gronau and D. London, Phys. Rev. Lett. 65 (1990) 3381; Y. Nir
and H. Quinn, Phys. Rev. Lett. 67 (1991) 541.

\bibitem{5} H. Fritzsch, D.D. Wu, and Z.Z. Xing, Phys. Lett. B328 (1994) 477;
Z.Z. Xing, Phys. Rev. D50 (1994) (in press).

\bibitem{6} J.P. Silva and L. Wolfenstein, Phys. Rev. D49 (1994) R1151.

\bibitem{7} M. Gronau, J.L. Rosner, and D. London, Phys. Rev. Lett. 73 (1994)
21;
O.F. Hern$\rm\acute{a}$ndez, D. London, M. Gronau, and J.L. Rosner,
Phys. Lett. B333 (1994) 500; M. Gronau, O.F. Hern$\rm\acute{a}$ndez,
D. London, and J.L. Rosner, Hep-Ph/9404283 (to appear in Phys. Rev. D).

\bibitem{8} Z.Z. Xing, LMU-13/94 (1994).

\bibitem{9} Particle Data Group, K. Hikasa et al., Phys. Rev. D45-II (1992) 1.

\bibitem{10} H. Fritzsch and P. Minkowski, Phys. Rep. 73 (1981) 67;
L.L. Chau, Phys. Rep. C178 (1983) 1.

\bibitem{11} A.J. Buras et al., Nucl. Phys. B375 (1992) 501; B370 (1992) 69.

\bibitem{12} A.J. Buras, J.M. G$\rm\acute{e}$rard, and R. R$\rm\ddot{u}$ckl,
Nucl. Phys. B268 (1986) 16; B. Blok and M. Shifman, Nucl. Phys. B389 (1993)
534.

\bibitem{13} M. Gronau and D. Wyler, Phys. Lett. B265 (1991) 172; I. Dunietz,
Phys. Lett. B270 (1991) 75.

\bibitem{14} A. Ali and D. London, CERN-Th.7248/94 (1994); J.L. Rosner, EFI
94-21 (1994).

\bibitem{15} CLEO Collaboration, M. Battle et al., Phys. Rev. Lett. 71 (1993)
3922.

\bibitem{16} M. Bauer, B. Stech, and M. Wirbel, Z. Phys. C34 (1987) 103.

\end{thebibliography}
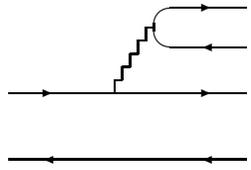
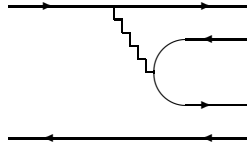
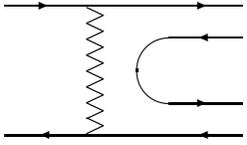
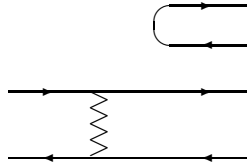
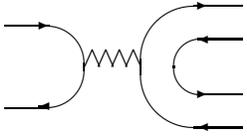
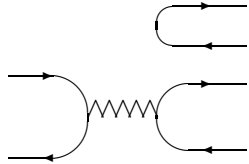
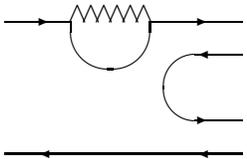
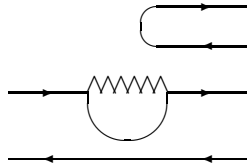
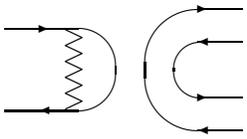
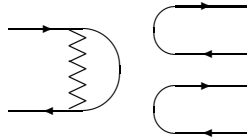
\end{document}